\documentclass[twocolumn,floats,superscriptaddress]{revtex4}
\usepackage{graphicx,epsfig}
\usepackage{times}
\usepackage{graphics,dcolumn,bm,float}
\usepackage{amssymb,amsmath,rotate,color}
\usepackage[normalem]{ulem}

\begin{document}

\title{Gate-controlled valley transport and Goos-H\"{a}nchen effect in monolayer WS$_2$}
\author{Hassan Ghadiri}
\affiliation{Department of Physics, North Tehran Branch, Islamic
Azad University, 16511-53311, Tehran, Iran}
\author{Alireza Saffarzadeh}\email{asaffarz@sfu.ca}
\affiliation{Department of Physics, Payame Noor University, P.O.
Box 19395-3697 Tehran, Iran} \affiliation{Department of Physics,
Simon Fraser University, Burnaby, British Columbia, Canada V5A
1S6}
\date{\today}

\begin{abstract}
Based on a Dirac-like Hamiltonian and coherent scattering
formalism, we study spin-valley transport and Goos-H\"{a}nchen
like (GHL) effect of transmitted and reflected electrons in a
gated monolayer WS$_2$. Our results show that the lateral shift of
spin-polarized electrons is strongly dependent on the width of the
gated region and can be positive or negative in both Klein
tunneling and classical motion regimes. The absolute values of the
lateral displacements at resonance positions can be considerably
enhanced when the incident angle of electrons is close to the
critical angle. In contrast to the time reversal symmetry for the
transmitted electrons, the GHL shift of the reflected beams is not
invariant under simultaneous interchange of spins and valleys,
indicating the lack of spin-valley symmetry induced by the tunable
potential barrier on WS$_2$ monolayer. Our findings provide
evidence for electrical control of valley filtering and valley
beam splitting by tuning the incident angle of electrons in
nanoelectronic devices based on monolayer transition metal
dichalcogenides.
\end{abstract}
\pacs{}

\maketitle

\section{Introduction}
Two-dimensional (2D) materials such as monolayers of transition
metal dichalcogenides (TMDs) have attracted considerable attention
due to their fascinating physical properties. In contrast to the
zero-gap graphene, they are direct gap semiconductors with sizable
bandgaps in the visible spectrum which make them promising
materials in electronic and optoelectronic applications such as
field-effect transistors and light-emitting diodes \cite{Wang,Jo}.
The layered TMDs such as MoS$_2$ and WS$_2$ monolayers possess
strong spin-orbit coupling due to the $d$ orbitals of heavy metal
atoms, causing a large spin splitting ($\sim$ 440 meV in WS$_2$
monolayers) in the valence bands \cite{ZhuYe,Salehi}. Moreover,
the absence of inversion symmetry in the crystal structure of
these materials allows valley polarization of carriers by
optically exciting electrons with a circularly polarized light
\cite{Yao,Xiao,Zeng,Mak,Sie,Ye}. The valleys occur at the
non-equivalent K and K$^\prime$ points in the hexagonal Brillouin
zone of TMD monolayers \cite{Xiao}. Manipulating both the spin and
valley degree of freedoms in such monolayers makes it possible to
design new generation of nanoelectronic devices for quantum
computing \cite{Zeng,Mak,Xiao,Ye}.

It was demonstrated that the electrons and holes at the band edges
of MoS$_2$ monolayers and other layered group-VI dichalcogenides
can be well described by massive Dirac fermions with strong
spin-valley coupling \cite{Xiao}. On the other hand, due to the
similarities between propagation of Dirac fermions and propagation
of electromagnetic waves in dielectrics, many optical phenomena
such as Brewster angles \cite{Wu}, collimation \cite{Park}, Bragg
reflection, electronic lenses \cite{Cheianov,Moghaddam} and
Goos-H\"{a}nchen (GH) shift \cite{Goos,Beenakker} have been found
in single-layer graphene. Among them, the GH effect refers to a
lateral displacement of a light beam when it is totally reflected
from a dielectric interface. In the past several years, most
studies on the electronic GH shift have been devoted to
graphene-based nanostructures such as graphene p-n junctions
\cite{Beenakker}, strained graphene \cite{Wu,Zhai}, graphene with
electric and magnetic barriers \cite{Ghosh,Agrawal}, and valley
\cite{Zhai} and spin \cite{Zhang} beam splitters in graphene.
Recently, the GH effect of electrons has also been studied in
silicene \cite{Azarova} and a p-n-p junction of MoS$_2$ monolayer
\cite{Sun}. It was found that the GH shift of Dirac fermions,
which has a magnitude of the order of Fermi wavelength
\cite{ChenJoO,ChenEPJB}, can be amplified by multiple total
internal reflections \cite{Sun,ChenJoO}. Moreover, the lateral
shifts of Dirac fermions in transmission through semiconductor
barriers \cite{ChenPLA}, graphene single barrier \cite{ChenEPJB},
double barrier \cite{Song}, and multiple barrier structures
\cite{ChenEPJB2} have also been reported. Such displacements,
called Goos-H\"{a}nchen like (GHL) shifts, occur in the partial
reflection regime and can be intensified by the transmission
resonances (Fabry-Perot resonances). However, it is still a
challenging problem to experimentally observe the GH or GHL shifts
of electrons due to the smallness of the shifts and also the
difficulty in producing a well-collimated electron beam
\cite{ChenJoO}. On the other hand, to design a beam splitter
effectively, the difference between lateral shifts of K and
K$^\prime$ valley electrons should be greater than the width of
incident electron beam which is about 100-1000 $\lambda_F$, with
$\lambda_F$ the Fermi wavelength of electrons \cite{Azarova,Song}.

In this paper we study the quantum GHL effect of spin polarized
electrons in transmission through a gated monolayer WS$_2$, acting
as a tunable potential barrier, as shown in Fig. 1(a), in two
cases of Klein tunneling and classical motion. We show that large
positive and negative GHL shift values can be obtained in the
transmitted and reflected electron beams. Moreover, the critical
angle for total reflection of electrons at the interface between
gated and ungated (normal) regions, is spin- and valley-dependent.
Accordingly, it is possible to choose incident angles, so that
spin-polarized electrons in one valley are allowed to propagate
through the gated region, whereas the electrons in the other
valley are blocked. We also show that the GHL effect can generate
transmitted and reflected valley-polarized beams with a separation
as large as the width of the incident beam, when a spin-polarized
electron with a proper choice of energy is incident on the
adjusted electrostatic barrier.

The paper is organized as follows. In section II, we introduce our
model and formalism for calculation of valley transport and GHL
shifts for both transmitted and reflected electron beams. It is
shown in section III for some limiting cases that our formalism
can also reproduce the results of lateral shifts of
\textit{totally} reflected beam in MoS$_2$ \cite{Sun} and graphene
single interfaces \cite{Beenakker}, and also \textit{partially}
reflected electrons from graphene barrier \cite{ChenEPJB}.
Numerical results and discussions for spin-valley transport and
lateral shifts in WS$_2$ monolayer, by tuning system parameters,
are presented in Sec. IV. A brief conclusion is given in Sec. V.

\section{Model and formalism}
To explore the effect of GHL shift in a TMD monolayer with a
tunable potential barrier, we consider a single-layer of WS$_2$ in
$x-y$ plane and apply a top gate voltage to the region II
($0<x<d$), while the regions I ($x<0$) and III ($x>d$) are kept at
zero electrostatic potential, as shown in Fig. 1. The top gate
voltage induces a potential barrier with height $V_G$ (Fig. 1(b)),
and turn the system into a n-p-n WS$_2$ junction. This means that
the electronic transport through the junction can be strongly
affected by the gate. We emphasize here that the gate voltage in
the proposed model is only utilized to shift the energy levels but
not to modulate the energy gap. In this normal/gated/normal WS$_2$
junction, the low-energy electrons near valleys K and K$^\prime$
can be described by the Dirac-like Hamiltonian \cite{Xiao}
\begin{eqnarray}\label{H}
\mathcal{H}&=&at(\tau
k_x\hat{\sigma}_x+k_y\hat{\sigma}_y)+\frac{\Delta}{2}\hat{\sigma}_z-
\lambda\tau\frac{\hat{\sigma}_z-1}{2}\hat{s}_z\nonumber\\
&&+V_{G}\Theta(x)\Theta(d-x)\ ,
\end{eqnarray}
where $a$ is the lattice constant, $t$ is the effective hopping
integral, and $\tau=1(-1)$ is the valley index corresponding to K
(K$^\prime$) point. $k_{x,y}$ denote the components of electron
wave vector measured from K and K$^\prime$, $\sigma_{x,y,z}$ are
the Pauli matrices spanning the conduction and valence states in
the two valleys, $\Delta$ is the band gap between valence and
conduction bands, and $2\lambda$ is the spin-splitting at the
valence band maximum due to the spin-orbit interaction.
$\hat{s}_z$ is the $z$ component of Pauli matrix for electron spin
and $\Theta(x)$ is the Heaviside step function. According to Eq.
(\ref{H}), the dispersion relation in the gated region II is given
by
\begin{equation}\label{Ek}
(2E-2V_G-\tau s_z\lambda)^2-(\Delta-\tau s_z\lambda)^2=(2atk')^2\
,
\end{equation}
where $k'=\sqrt{k'^2_x+k^2_y}$ and $s_z=\pm 1$ which stands for
spin up ($\uparrow$) and spin down ($\downarrow$). In the regions
I and III, however, the relation can be obtained by setting
$V_G=0$ and replacing $k'$ with $k=\sqrt{k^2_x+k^2_y}$.
\begin{figure}
\center\includegraphics[width=0.95\linewidth]{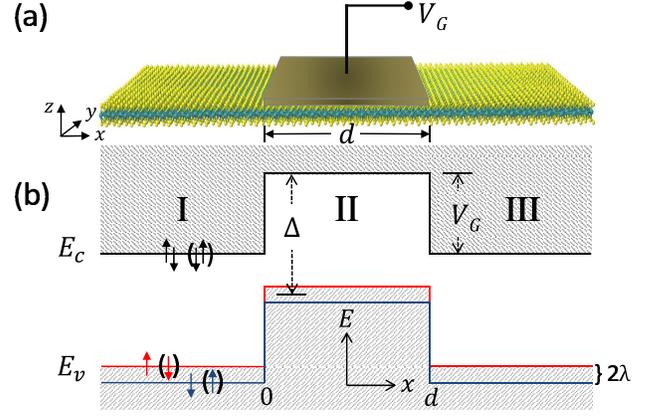}
\caption{(color online) Schematic model for the gated monolayer
WS$_2$ in which the gated region acts as a tunable potential
barrier. (b) Energy profile of the model shown in (a) for an
incident electron with spin $s_z (\bar{s}_z)$ in valley K
(K$^\prime$) and a typical value $V_G<\Delta$. $E_c$ and $E_v$
denote the energy of conduction band minimum and valence band
maximum, respectively. The red and blue lines show the
spin-dependent energy levels at the top of the valence band, due
to the spin-orbit coupling. Note that the spin splitting at
different valleys are opposite due to the time-reversal symmetry.}
\end{figure}

Now we consider the propagation of spin-polarized electrons in the
WS$_2$ junction which can be scattered by the potential barrier,
induced by the gate voltage $V_G$. It is assumed that the
spin-polarized electrons are injected by a magnetic contact, such
as a ferromagnetic semiconductor \cite{Ye,Saffar}, into the left
normal region. For simplicity, the source of spin injection is not
included in the model. We further assume that the dimension of
monolayer along the $y$-axis is infinitely long. Therefore, due to
the translational invariance in the $y$ direction, $k_y$ is
conserved, and the electron wave function in each region can be
described as
\begin{eqnarray}\label{psi1}
\psi_{1}&=&\frac{1}{B_1}\left(\begin{array}{c}
A_1\\ \tau atk\mathrm{e}^{i\tau\varphi}\\
\end{array}\right)\mathrm{e}^{i(k_xx+k_yy)}\nonumber\\
&&+\frac{r_{s_z}^\tau}{B_1}\left(\begin{array}{c}
A_1\\ -\tau atk\mathrm{e}^{-i\tau\varphi}\\
\end{array}\right)\mathrm{e}^{i(-k_xx+k_yy)}\  ,
\end{eqnarray}
\begin{eqnarray}\label{psi2}
\psi_{2}&=&\frac{\alpha}{B_2}\left(\begin{array}{c}
A_2\\ \tau atk'\mathrm{e}^{i\tau\varphi'}\\
\end{array}\right)\mathrm{e}^{i(k'_xx+k_yy)}\nonumber\\
&&+\frac{\beta}{B_2}\left(\begin{array}{c}
A_2\\ -\tau atk'\mathrm{e}^{-i\tau\varphi'}\\
\end{array}\right)\mathrm{e}^{i(-k'_xx+k_yy)}\  ,
\end{eqnarray}
\begin{eqnarray}\label{psi3}
\psi_{3}&=&\frac{t_{s_z}^\tau}{B_3}\left(\begin{array}{c}
A_3\\ \tau atk\mathrm{e}^{i\tau\varphi}\\
\end{array}\right)\mathrm{e}^{i(k_xx+k_yy)}\  ,
\end{eqnarray}
where $\varphi=\mathrm{tan}^{-1}(k_y/k_x)$ is the incident angle
of electrons on the barrier, $r_{s_z}^\tau$ is the reflection
amplitude, $\varphi'=\mathrm{tan}^{-1}(k_y/k'_x)$ is the angle of
refraction, and $t_{s_z}^\tau$ is the transmission amplitude of
electrons. The coefficients $A_i$ and $B_i$ ($i=1,2,3$) in Eq.
(\ref{psi1}-\ref{psi3}) are expressed as $A_1=A_3=E-\tau
s_z\lambda+\Delta/2$, $A_2=A_1-V_G$,
$B_1=B_3=\sqrt{A_1^2+(atk)^2}$, and $B_2=\sqrt{A_2^2+(atk')^2}$.

The total internal reflection of electrons occurs if the incident
electron waves strike the potential barrier at a critical angle
$\varphi_c$ which is given by
\begin{equation}\label{phic}
\varphi_c=\arcsin\sqrt{\frac{(2E-2V_G-\tau
s_z\lambda)^2-(\Delta-\tau s_z\lambda)^2}{(2E-\tau
s_z\lambda)^2-(\Delta-\tau s_z\lambda)^2}}\ .
\end{equation}
When the angle of incidence $\varphi$ is greater than the critical
angle, the electron wave vector in the electrostatic barrier,
$k'_x=\sqrt{k'^2-k^2_y}$, becomes imaginary which indicates an
evanescent wave in the gated region, and hence, the electron beam
will experience a total reflection from the electrostatic barrier.
In the case of $\varphi<\varphi_c$, however, the electron wave is
partially reflected and partially transmitted. The coefficients
$r^\tau_{s_z}$, $\alpha$, $\beta$, and $t_{s_z}^\tau$ can be
obtained by applying the continuity of the wave functions at the
two interfaces, $x=0$ and $x=d$ shown in Fig. 1. The transmission
$t^\tau_{s_z}$ and reflection $r^\tau_{s_z}$ coefficients for
electron waves are expressed as
\begin{equation}\label{tr}
t^\tau_{s_z}=\frac{2ss'\sqrt{\frac{F_1}{F_2}}\cos\varphi\cos\varphi'
\mathrm{e}^{-ik_xd}}{2ss'\sqrt{\frac{F_1}{F_2}}~\cos\varphi\cos\varphi'\cos(k'_xd)-iD}\
,
\end{equation}

\begin{equation}\label{r}
r^\tau_{s_z}=\frac{[-2ss'\sqrt{\frac{F_1}{F_2}}~\mathrm{e}^{i\tau\varphi}\sin(\tau\varphi')+i(1-\frac{F_1}{F_2}\mathrm{e}^{2i\tau\varphi})]
\sin(k'_xd)}{2ss'\sqrt{\frac{F_1}{F_2}}~\cos\varphi\cos\varphi'\cos(k'_xd)-iD}\
,
\end{equation}
where
$D=[(1+\frac{F_1}{F_2})-2ss'\sqrt{\frac{F_1}{F_2}}\sin\varphi\sin\varphi']
\sin(k'_xd)$, $F_1=A_2/A_1$, $F_2=(E-V_G-\Delta/2)/(E-\Delta/2)$,
$s=\mathrm{sgn}(A_1)$, and $s'=\mathrm{sgn}(A_2)$.

To study the GHL effect, we consider an incident electron in the
form of a Gaussian wave packet of width $\Delta_{k_{y}}$ and
energy $E$,
\begin{eqnarray}\label{psiin}
\psi_{in}(x,y)&=&\int_{-\infty}^{+\infty}dk_yf(k_y-k_{y_0})\frac{1}{B_1}\nonumber\\
&&\times\left(\begin{array}{c}
A_1\\
\tau
atk\mathrm{e}^{i\tau\varphi(k_y)}\end{array}\right)\mathrm{e}^{i[k_x(k_y)x+k_yy]}\
,
\end{eqnarray}
where the Gaussian
$f(k_y-k_{y_0})=\mathrm{e}^{-(k_y-k_{y_0})^2/2\Delta_{k_y}^2}$
shows the angular distribution of electron beam around the central
incidence angle $\varphi_0=\arcsin{(k_{y_0}/k)}$. Therefore, the
transmitted electron beam can be given as
\begin{eqnarray}\label{psitr}
\psi_{tr}(x,y)&=&\int_{-\infty}^{+\infty}dk_yf(k_y-k_{y_0})\frac{t^{\tau}_{s_z}(k_y)}{B_3}\nonumber\\
&&\times\left(\begin{array}{c}
A_3\\
\tau
atk\mathrm{e}^{i\tau\varphi(k_y)}\end{array}\right)\mathrm{e}^{i[k_x(k_y)x+k_yy]}\
.
\end{eqnarray}

In the case of $\Delta_{k_y}\ll k$, we can expand $\varphi(k_y)$
and $k_x(k_y)$ to first order around $k_{y_0}$, substitute in Eqs.
(\ref{psiin}) and (\ref{psitr}), and evaluate the integrals, to
obtain the spatial form of the incident and transmitted electron
beams. In this regard, the incident beam is given by
\begin{eqnarray}\label{psiin1}
\psi_{in}(x,y)&=&\frac{\sqrt{2\pi}\Delta_{k_{y}}}{B_1}\left(\begin{array}{c}
A_1~\mathrm{e}^{-(y-\bar{y}^{in}_{+})^2\Delta^2_{k_{y}}/2}\\
\tau
atk\mathrm{e}^{-(y-\bar{y}^{in}_{-})^2\Delta^2_{k_{y}}/2}~\mathrm{e}^{i\tau\varphi(k_{y_0})}\end{array}\right)\nonumber\\
&&\times~\mathrm{e}^{i[k_x(k_{y_0})x+k_{y_0}y]}\  ,
\end{eqnarray}
where $\bar{y}^{in}_{+}=-\dot{k}_x(k_{y_0})x$, and
$\bar{y}^{in}_{-}=-\dot{k}_x(k_{y_0})x-\tau\dot{\varphi}(k_{y_0})$.
Here, the dot indicates the derivative with respect to $k_y$.
Therefore, the two components of $\psi_{in}$ represent two
Gaussians of the same width, centered at the two different mean
coordinates $\bar{y}^{in}_{+}$ and $\bar{y}^{in}_{-}$ along the
y-axis.

Using a similar analysis, the transmitted electron beam can be
expressed as
\begin{eqnarray}\label{psiin2}
\psi_{tr}(x,y)&=&\frac{\sqrt{2\pi}\Delta_{k_{y}}}{B_3}~\mathrm{e}^{i[k_x(k_{y_0})x+k_{y_0}y]}|t^\tau_{s_z}(k_{y_0})|\nonumber\\
&\times&\left(\begin{array}{c}
A_3~\mathrm{e}^{-\frac{1}{2}(y-\bar{y}^{tr}_{+})^2\Delta^2_{k_{y}}}\\
\tau
atk\mathrm{e}^{-\frac{1}{2}(y-\bar{y}^{tr}_{-})^2\Delta^2_{k_{y}}}\mathrm{e}^{i[\tau\varphi(k_{y_0})+\Phi_{t^\tau_{s_z}}(k_{y_0})]}
\end{array}\right)\  ,
\end{eqnarray}
where
$\bar{y}^{tr}_{+}=-\dot{k}_x(k_{y_0})x-\dot{\Phi}_{t^\tau_{s_z}}(k_{y_0})$
and
$\bar{y}^{tr}_{-}=-\dot{k}_x(k_{y_0})x-\dot{\Phi}_{t^\tau_{s_z}}(k_{y_0})-\tau\dot{\varphi}(k_{y_0})$.
Note that $\Phi_{t^\tau_{s_z}}$ in Eq. (\ref{psiin2}) and in the
mean coordinates $\bar{y}^{tr}_{\pm}$, denotes the phase of
transmission coefficient $t^\tau_{s_z}$. Accordingly, the GHL
shifts of the upper and lower components of the transmitted
electrons are given by
\begin{eqnarray}\label{sigp}
\sigma_{\pm}&=&\bar{y}^{tr}_{\pm}|_{x=d}-\bar{y}^{in}_{\pm}|_{x=0}\nonumber\\
&=&-\dot{\Phi}_{t^\tau_{s_z}}(k_{y_0})-\dot{k}_x(k_{y_0})d\  .
\end{eqnarray}
Since the two components have equal shift values, the average GHL
shift of the transmitted beam can be expressed as
\begin{equation}\label{GHLt}
\sigma^\tau_{tr,s_z}=-\dot{\Phi}_{t^\tau_{s_z}}(k_{y_0})-\dot{k}_x(k_{y_0})d\
.
\end{equation}
Applying similar considerations to the lateral shift of the
reflected beam, we find that
$\sigma_+=-\dot{\Phi}_{r^\tau_{s_z}}(k_{y_0})$ and
$\sigma_-=\sigma_++2\tau\dot{\varphi}(k_{y_0})$. By considering
the probability of upper and lower components of the beams, the
average shift of reflected electrons will be
\begin{eqnarray}\label{GHLr}
\sigma^\tau_{re,s_z}&=&\frac{1}{{B_1}^2}[{A_1}^2\sigma_++(atk)^2\sigma_-]\nonumber\\
&=&-\dot{\Phi}_{r^\tau_{s_z}}(k_{y_0})+\frac{2\tau
a^2t^2k^2}{{B_1}^2}\dot{\varphi}(k_{y_0}).
\end{eqnarray}

In the case of transmitted beam, we need to calculate
$\Phi_{t_{s_z}}^\tau$ from Eq. (\ref{tr}), and substitute into Eq.
(\ref{GHLt}). Thus, the lateral GHL shift of transmitted electrons
through the tunable electrostatic potential can be obtained by
\begin{equation}\label{GHLt-end}
\sigma^{\tau}_{tr,s_z}=\frac{[(8+2\frac{k^2_0}{k^2_x}+2\frac{k^2_0}{k'^2_x})\frac{\sin(2k'_xd)}{2k'_xd}-2\frac{k^2_0}{k'^2_x}]d\tan\varphi}
{4\cos^2(k'_xd)+\frac{k_0^4}{k^2_xk'^2_x}\sin^2(k'_xd)}\ ,
\end{equation}
where $k^2_0=-ss'\gamma kk'+2k^2_y$ and
$\gamma=\sqrt{F_2/F_1}(1+F_1/F_2)$.

On the other hand, by computing $\Phi_{r^\tau_{s_z}}$ from Eq.
(\ref{r}) and substituting into Eq. (\ref{GHLr}), the GHL shift
for the reflected wave can be written as
\begin{equation}\label{GHLr-end}
\sigma^\tau_{re,s_z}=\sigma^\tau_{tr,s_z}+\frac{2\tau}{k_x}\left[\frac{C_1+C_2-C_3}{B^2_1(1+C_0)}\right]\
,
\end{equation}
where
\begin{eqnarray}\label{C0}
C_0&=&\left[\frac{F_1}{F_2}+4\frac{k^2_y}{k'^2}
-2\frac{(k_x^2-k_y^2)}{k^2}\right]\frac{F_1}{F_2}\nonumber\\
&&-4ss'\frac{k^2_y}{kk'}(1+\frac{F_1}{F_2})\sqrt{\frac{F_1}{F_2}}\
,
\end{eqnarray}
\begin{equation}\label{C1}
C_1=a^2t^2k^2-A^2_1\frac{F^2_1}{F^2_2}+(a^2t^2k^2-A^2_1)\left[\frac{2k^2_y}{k'^2}
+\frac{k^2_y-k^2_x}{k^2}\right]\frac{F_1}{F_2}\ ,
\end{equation}
\begin{equation}\label{C2}
C_2=ss'\left[k^2_x(a^2t^2k^2+A^2_1)-k^2_y(a^2t^2k^2-3A^2_1)\right]\frac{1}{kk'}\sqrt{\frac{F^3_1}{F^3_2}}
\ ,
\end{equation}
\begin{equation}\label{C3}
C_3=ss'\left[k^2_x(a^2t^2k^2+A^2_1)+k^2_y(3a^2t^2k^2-A^2_1)\right]\frac{1}{kk'}\sqrt{\frac{F_1}{F_2}}
\ .
\end{equation}

It is important to note that the GHL shift for the transmitted
wave is dependent on the product of $s_z$ and $\tau$, as can be
seen from the expressions, $F_1$, $F_2$, $s$, $s'$, and Eq.
(\ref{Ek}). In other words, the GHL shift for the transmitted wave
has a $s_z\cdot\tau$ (spin-valley) symmetry \cite{PNAS}, whereas
such a symmetry cannot be seen in the GHL shift of the reflected
wave, due to the presence of the second term in Eq.
(\ref{GHLr-end}). In fact, the reflected beam is not only
dependent on the product of $\tau$ and $s_z$, but also on valley
index $\tau$, separately, indicating that the reflected electron
does not remain invariant under simultaneous interchange of spins
and valleys in this system.

\section{Limiting cases in GHL shift}
Now we consider some limiting cases to show how the present
formalism is also able to reproduce the analytic results of
previous studies in MoS$_2$ and graphene single interfaces
\cite{Sun,Beenakker}, and also graphene barrier \cite{ChenEPJB}.
\begin{figure}
\center\includegraphics[width=0.95\linewidth]{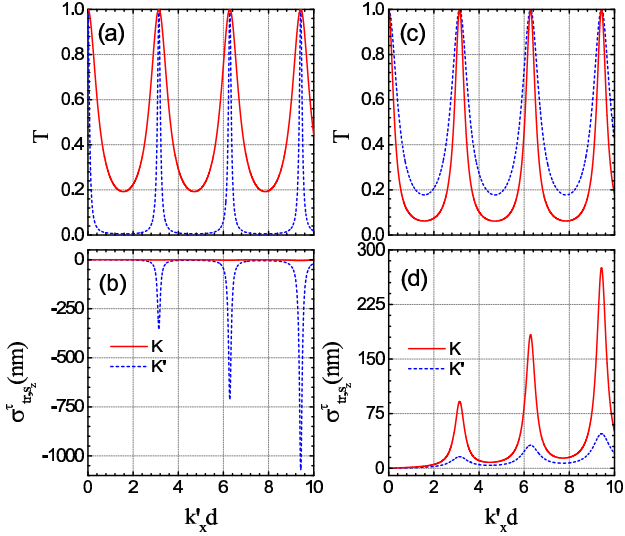}
\caption{(color online) Dependence of transmission probability and
GHL shift of the transmitted electrons with $s_z=1$ on the width
$d$ of the gated region in (a,b) Klein tunneling and (c,d)
classical motion, where $V_G$=3 eV. The other parameters are:
(a,b) $E=1.75$ eV, $\varphi=19.5^\circ$, and
$\varphi_c=50.10^\circ$ in K-valley, $\varphi_c=20.51^\circ$ in
K$'$-valley, (c,d) $E=5.0$ eV, $\varphi=20.5^\circ$, and
$\varphi_c=20.88^\circ$ in K-valley, $\varphi_c=21.73^\circ$ in
K$'$-valley. Note that $d$ is normalized with $k'_x$.}
\end{figure}

\subsection{single interface on TMD monolayers}
As already mentioned, for incident angles greater than the
critical angle, $\varphi
>\varphi_c$, the wave vector $k'_x$ becomes imaginary and hence,
the total reflection occurs. In such a case, Eq. (\ref{GHLr-end})
which is given for $\varphi <\varphi_c$ is also valid for $\varphi
>\varphi_c$, provided that we make the substitution $k'_x\rightarrow i\kappa$.
After this substitution and taking the limit $d\rightarrow
\infty$, the second term in Eq. (\ref{GHLr-end}) does not change,
but the first term becomes
\begin{equation}\label{second-term}
\frac{[8k^2_x\kappa^2+2k_0^2(\kappa^2-k_x^2)]\tan
\varphi}{\kappa(4k_x^2\kappa^2+k_0^4)}\ .
\end{equation}
Therefore, the GH shift of the totally reflected beam from a step
potential (single interface) applied on a monolayer of TMD, can be
easily calculated by replacing the first term of Eq.
(\ref{GHLr-end}) by (\ref{second-term}).

In Ref. (\onlinecite{Sun}), the probabilities of upper and lower
components of the reflected electrons have not been included in
calculations and the average GH shift was simply computed by
$\sigma^\tau_{re,s_z}=\frac{1}{2}(\sigma_++\sigma_-)$. On this
basis, the second term in Eq. (\ref{GHLr-end}) becomes
$\frac{\tau}{k_x}[1-\frac{F^2_1}{F^2_2}-2ss'\sqrt{\frac{F_1}{F_2}}(1-\frac{F_1}{F_2})\frac{k}{k'}]/(1+C_0)$,
and finally the GH shift of the totally reflected beam can be
simplified as
\begin{equation}\label{sigma-interface}
\sigma^{\tau}_{{re}_{total},s_z}=\frac{\frac{\tau}{k_x}[(\tau\kappa+
k_y)^2-k'^2\frac{F_1}{F_2}]+\frac{2ss'k'}{\tau\kappa}(\kappa+\tau
k_y)\sqrt{\frac{F_1}{F_2}}
\cos\varphi}{(\tau\kappa+k_y)^2+k'^2\frac{F_1}{F_2}-2ss'k'(\tau\kappa+
k_y)\sqrt{\frac{F_1}{F_2}}\sin\varphi}\ .
\end{equation}
This equation is exactly the same as Eq. (8) given in Ref.
\onlinecite{Sun} for MoS$_2$ monolayers.

\subsection{graphene barrier}
By setting $\lambda=0$ and $\Delta=0$ in Eqs. (\ref{H}),
(\ref{GHLt-end}), and (\ref{GHLr-end}), the Hamiltonian of gapless
graphene with a tunable potential barrier and the corresponding
GHL shifts for transmitted and reflected beams can be obtained,
respectively \cite{ChenEPJB}. In this regard, the lateral shifts
of the reflected and transmitted electron beams in graphene are
the same and expressed as
\begin{equation}\label{GHLtr-graphene}
\sigma_{re}=\sigma_{tr}=\frac{[(2+\frac{k'^2_0}{k^2_x}+\frac{k'^2_0}{k'^2_x})\frac{\sin(2k'_xd)}{2k'_xd}-\frac{k'^2_0}{k'^2_x}]d\tan\varphi}
{\cos^2(k'_xd)+\frac{k'^4_0}{k^2_xk'^2_x}\sin^2(k'_xd)}\ ,
\end{equation}
where $k_0'^2=-ss'kk'+k_y^2$. Note that in the case of $\lambda=0$
and $\Delta\neq 0$ (gapped graphene), we obtain
$\sigma_{re}\neq\sigma_{tr}$.
\begin{figure}
\center\includegraphics[width=0.95\linewidth]{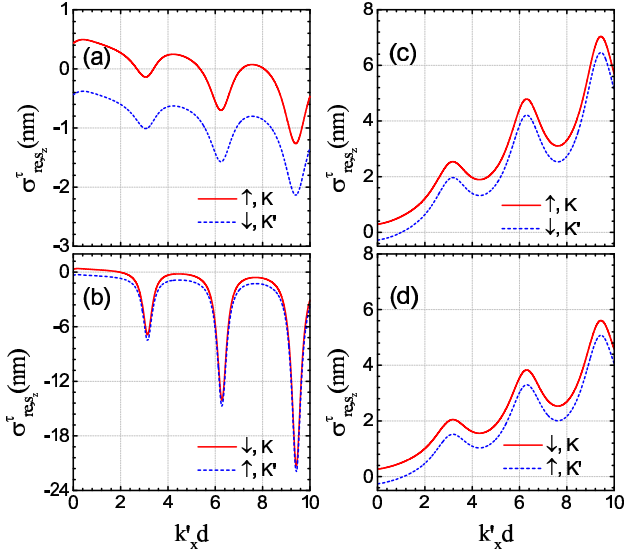}
\caption{(color online) Dependence of GHL shift in the reflected
beam on the width $d$ of the gated region in (a,b) Klein tunneling
and (c,d) classical motion, where for all plots $\varphi=10^\circ$
and $V_G=3$ eV. The other parameters are: (a) $E=1.75$ eV and
$\varphi_c=50.10^\circ$, (b) $E=1.75$ eV and
$\varphi_c=20.51^\circ$, (c) $E=4.45$ eV and
$\varphi_c=14.75^\circ$, (d) $E=4.45$ eV and
$\varphi_c=15.55^\circ$. Note that $d$ is normalized with $k'_x$.}
\end{figure}

Eq. (\ref{GHLtr-graphene}) at $ss'=\pm1$ is the same as Eqs. (11)
and (19) in Ref. \onlinecite{ChenEPJB}, respectively. Moreover, in
the case of $ss'=-1$, substituting $k_x'\rightarrow i\kappa$ for
$\varphi>\varphi_c$ into Eq. (\ref{GHLtr-graphene}), and taking
the limit $d\rightarrow\infty$, one can obtain after some
calculation
\begin{equation}\label{sig1}
\sigma=\frac{k_y^2-kk'}{\kappa k_xk_y}\  .
\end{equation}
Finally, using Eq. (\ref{Ek}), we obtain
\begin{equation}\label{sig2}
\sigma=\frac{\sin^2\varphi+1-V_G/E}{\kappa\sin\varphi\cos\varphi}\
.
\end{equation}
This equation describes the GH shift of electrons in total
reflection from a step potential (single interface) on graphene.
It is interesting to note that this expression is exactly the same
as Eq. (11) given in Ref. \onlinecite{Beenakker}. Therefore, our
present formalism in limiting cases is able to reproduce the
previous results of GH shifts for single interfaces on graphene
\cite{error} and TMD monolayers \cite{Sun}, and also GHL shift on
graphene barrier \cite{ChenEPJB}.

\section{results and discussion}
We present our numerical result for gated monolayer WS$_2$ as a
family member of layered TMD materials, using the parameters
$t$=1.37 eV, $\Delta$=1.79 eV, 2$\lambda$=0.43 eV, and $a$=3.197
{\AA} \cite{Xiao}. From the dispersion relation, Eq. (\ref{Ek}),
the energy regions for electrons in propagating mode are given by
$\frac{\Delta}{2}<E<V_G-\frac{\Delta}{2}+\tau s_z\lambda$ and
$E>V_G+\frac{\Delta}{2}$ associated with the Klein tunneling
effect \cite{Klein,BeenakkerRMP,Katsnelson,ChenEPJB} ($ss'=-1$)
and the classical motion ($ss'=1$), respectively. In both regimes,
transmission probability, $T=|t^{\tau}_{s_z}|^2$, and lateral
shift $\sigma^{\tau}_{tr,s_z}$ of electrons with energy value $E$
and a given $V_G$, are strongly dependent on the width $d$ of the
gated region. Therefore, to compare these quantities in the two
energy regions, we show in Fig. 2 the transmission probabilities
and the GHL shifts of a transmitted electron beam with spin up
($s_z=1$) through the gated monolayer, as the barrier width $d$
(normalized by $k'_x$) is increased. At the resonance conditions
$k'_{x}d=n\pi (n=0,\pm1,\pm2,\cdots)$, the potential barrier in
both energy regions becomes fully transparent, $T=1$, and the
maximum absolute values of the lateral shifts are given by
\begin{equation}\label{max}
\sigma^{\tau}_{tr,s_z}|_{k'_xd=n\pi}=\frac{k^2_{0}d\tan\varphi}{2k'^2_x}\
.
\end{equation}

Note that only the first three maxima are shown in Fig. 2(b) and
(d). In the Klein tunneling regime (Fig. 2(a) and (b)), the
transmission probabilities and the lateral shifts are strongly
dependent on the valleys K and K$'$, whereas such a
valley-dependent transport is weaker in the classical motion as
shown in Fig. 2(c) and (d). Although the angle of incidence is
closer to the critical angles in the classical regime compared to
the Klein tunneling regime, the GHL effect in the Klein tunneling
can more significantly separate the K and K$'$ valley electron
beams compared to the classical motion. For instance, the
difference between the two lateral displacements at the third
maximum absolute value is $\sim 1065$ nm in the Klein tunneling,
while it is $\sim 227$ nm in the classical regime.

From Eq. (\ref{GHLt-end}), it is clear that the lateral shift is
dependent on the angle of incidence. Moreover, $k'_x$ is zero at
critical angle $\varphi_c$. Therefore, at incident angles
$\varphi$ close to $\varphi_c$, the maximum absolute values of the
shifts given by Eq. (\ref{max}), can increase rapidly so that the
spatial separation between K and K$^\prime$ electron waves in both
regimes can exceed the width of incident beam. It should be
mentioned that due to the presence of $s_z\cdot\tau$ symmetry in
Eq. (\ref{GHLt-end}), the results given in Fig. 2 with $s_z=+1$
and $\tau=-1$ $(\tau=+1)$ are the same as those with $s_z=-1$ and
$\tau=+1$ $(\tau=-1)$ (not shown).
\begin{figure}
\center\includegraphics[width=0.65\linewidth]{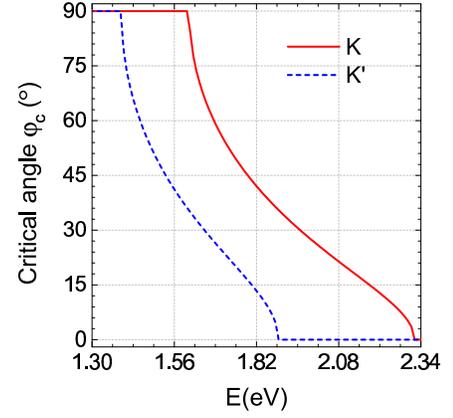}
\caption{(color online) Dependence of critical angles on incident
energy for electrons with $s_z=1$ in valleys K and K$'$ at $V_G$=3
eV.}
\end{figure}

We now study the effect of gated region on the reflected electron
beam with some parameters different with those in Fig. 2. The GHL
shifts as a function of normalized barrier width $k'_xd$ in the
both regimes are shown in Fig. 3. In the Klein tunneling effect
(Fig. 3(a) and (b)), the reflected K valley electrons exhibit
positive or negative values of the lateral shifts, depending on
the $d$ values, whereas the GHL shifts of electrons belonging to
the K$'$ valley are only negative. In the classical motion regime
(Figs. 3(c) and (d)), however, the reflected K$'$ valley electrons
can take positive or negative shift values, whereas the values are
solely positive for the reflected beam belonging to the K-valley.
The peaks in $\sigma^{\tau}_{re,s_z}$ correspond to the periodical
occurrence of transmission resonances, as can be seen in the
transmitted beam (see Fig. 2). Nevertheless, the lateral shift
values at the resonance positions for the reflected electron beams
are physically meaningless which is due to the fact that the
reflection probability, $R=1-T$, at these positions is zero. In
the vicinity of the resonances, the lateral shift of the reflected
beam (similar to that of the transmitted beam) can increase
rapidly when the incident angle approaches the critical angles.
\begin{figure}
\center\includegraphics[width=0.95\linewidth]{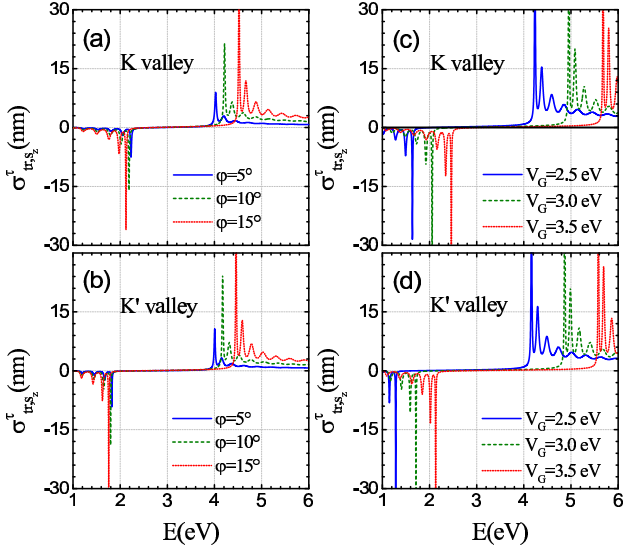}
\caption{(color online) The GHL shift of the transmitted K and
K$'$ electrons as a function of incident energy at (a,b) different
incident angles and (c,d) different gate voltages. In (a,b)
$V_G=3.0$ eV, and (c,d) $\varphi=20^\circ$. The width $d$ of the
gated region is fixed at 4 nm.}
\end{figure}

As already mentioned, the $s_z\cdot\tau$ symmetry seen in the
$\sigma^{\tau}_{tr,s_z}$ does not exist in
$\sigma^{\tau}_{re,s_z}$. Therefore, the values of lateral shifts
for reflected K-valley electrons with spin $s_z$ are different
from those for K$'$-valley electrons with spin $\bar{s}_z$. As a
result, the reflected electron beams from such electrostatic
barriers are fully spin- and valley-polarized. However, as can be
seen from Fig. 3, the lateral shift difference for the reflected
electrons with the same value of $s_z\cdot\tau$ (in both regimes)
is not enough to be detected experimentally, while it is
detectable for electrons with different values of $s_z\cdot\tau$,
when the incident angle is sufficiently close to $\varphi_c$ (not
shown).

From Eq. (\ref{phic}) it is clear that the critical angle
$\varphi_c$ for incident spin-polarized electrons with energy $E$
is valley-dependent. Fig. 4 shows that how this angle for an
incident electron with $s_z=+1$ and $\tau=\pm 1$ changes with
energy. We can see that the values of $\varphi_c$ within a
specific energy window are quite different for the two valleys.
This suggests that the incident angle $\varphi$ can be tuned so
that only electrons from a specific valley traverse through the
electrostatic barrier, while the electrons from the other valley
are fully reflected and blocked. For instance, for electrons with
$s_z=+1$ and incident energy $E=1.75$ eV, the critical angles in
two valleys are $\varphi_c({\tau=1})=50^\circ$ and
$\varphi_c(\tau=-1)=20^\circ$ (see Fig. 4). Therefore, if the
incident angle $\varphi$ ranges between $20^\circ$ and $50^\circ$,
the K$'$ valley electrons are backscattered, while the electrons
belonging to the K-valley are transmitted, due to the different
trajectories imposed by the gate voltage for each valley,
indicating a valley filter with wide tunability.

The energy dependence of GHL shift for the transmitted K and K$'$
valley electrons with $s_z=1$ through the gated region with $d=4$
nm is shown in Fig. 5. The lateral shifts of electrons at various
angles of incidence (Figs. 5(a) and (b)) demonstrate successive
peaks with different absolute values, corresponding to the
transmission resonances, in both Klein tunneling and classical
motion regimes with positive and negative lateral shifts,
respectively. The almost zero lateral shift region between Klein
tunneling and classical motion is the transmission gap
\cite{ChenAPL2009} and is given by $V_G+\frac{1}{2}\tau
s_z\lambda-\frac{1}{2}\sqrt{4a^2t^2k^2_y+(\Delta-\tau
s_z\lambda)^2}<E<V_G+\frac{1}{2}\tau
s_z\lambda+\frac{1}{2}\sqrt{4a^2t^2k^2_y+(\Delta-\tau
s_z\lambda)^2}$, which includes the energy gap as well.

Clearly the transmission gap can be wider and the absolute values
of the lateral shifts, as well as their difference belonging to
different valleys, can be enhanced by increasing the angle of
incidence. Moreover, by comparing the energies of maximum absolute
values of GHL shifts for K and K$'$ electrons, we can see that the
valley separation in the case of Klein tunneling is stronger than
that in the classical motion \cite{Ye}. Figs. 5(c) and (d)
illustrate the effect of various gate voltages on GHL shift of the
transmitted K and K$'$ electrons, when the incident angle is
20$^\circ$. Here, one can see that the transmission gap also
increases with increasing the external gate. In addition, with
increasing the gate voltage, the whole spectrum of the lateral
shifts moves towards higher energies and the maximum absolute
values of the shifts as well as their differences, increase.

Note that due to the time reversal symmetry, the lateral shifts of
incident electrons with spin $s_z=-1$, remain unchanged, if we
interchange the valleys in the transmitted beam, while the lateral
shifts will be somewhat different by interchanging the valleys in
the reflected electron beams (not shown) due to the absence of
$s_z\cdot\tau$ symmetry, as discussed previously.
\begin{figure}
\center\includegraphics[width=0.7\linewidth]{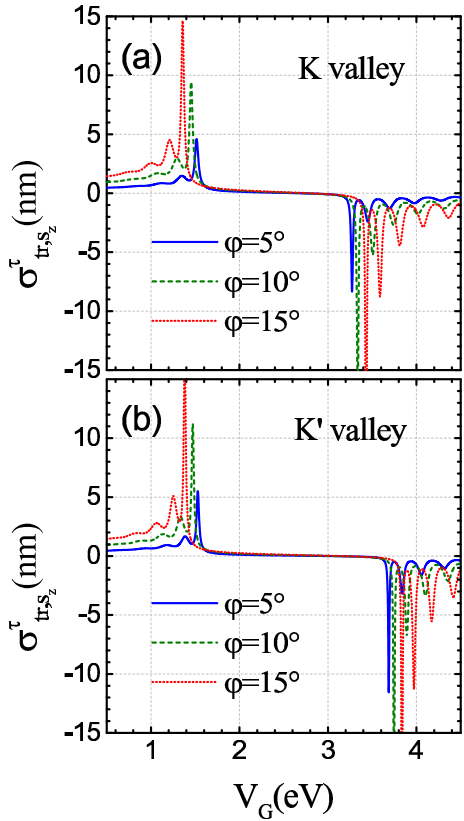}
\caption{(color online) Gate voltage dependence of GHL shift for
transmitted K and K$'$ electrons with $s_z=1$ at three different
incident angles. The incident energy $E$ and the width $d$ of the
gated region are fixed at 2.5 eV and 4 nm, respectively.}
\end{figure}

In order to explore the influence of a continues range of gate
voltages on the GHL shift, in Fig. 6 we show the calculated
results for transmitted K and K$'$ electrons with spin $s_z=1$ and
energy $E=2.5$ eV, when the external gate is varied from 0.5 to
4.5 eV. The lateral shifts of both K and K$'$ electrons are
positive at low gate values, due to the classical motion of
electrons which satisfy the condition $E>V_G+\frac{\Delta}{2}$ for
any incident angle $\varphi <\varphi_c$. However, at higher gate
voltages, the incident electrons experience the Klein tunneling
effect, and hence, the lateral shifts become negative. Again,
there is a gap-like feature in the lateral shift values between
the classical motion and Klein tunneling regimes, corresponding to
the transmission gap. As the incident angle increases, the maximum
absolute values of the lateral shifts as well as their differences
increase and the gap-like region becomes broader.

Comparing the gate voltages at which the absolute values of the
lateral shifts in the two valleys are maximum, we can see that at
some gate values the shift of K$'$ valley electrons in the Klein
tunneling is almost zero, while the shift of K valley electrons is
maximum. As already mentioned, these maximum values correspond to
the transmission resonances and demonstrate the valley splitting
of charge carriers in WS$_2$ monolayers, which can be generated
and controlled by gate voltage \cite{Ye,Mak}.

\section{conclusion}
In conclusion, the spin-valley transport and the GHL effect of the
transmitted and reflected electrons in a WS$_2$ monolayer with a
tunable electrostatic barrier have been theoretically studied in
both Klein tunneling and classical motion regimes. Interestingly,
it was found that the GHL shift of the reflected electrons does
not remain invariant under simultaneous interchange of spins and
valleys. Therefore, the spin-valley symmetry in the lateral shifts
of the reflected electrons from the barrier is different compared
to that in the transmitted electrons. The lateral displacement of
electron beams can be positive or negative dependent on the width
of the gated region and/or incident energy. Our findings show that
a valley filter and valley beam splitter can be achieved by tuning
the incident angle of electrons and external gate voltage in
WS$_2$ monolayers. These features indicate the possibility of
manipulating valley index in the TMD monolayers which can be
utilized for quantum information applications \cite{Ye}.

\section*{Acknowledgement}
This work is partially supported by Iran Science Elites
Federation.


\begin{thebibliography}{99}

\bibitem{Wang}
{Q. H. Wang, K. Kalantar-Zadeh, A. Kis, J. N. Coleman, and M. S.
Strano, Nat. Nanotech. {\bf 7}, 699 (2012).}

\bibitem{Jo}
{S. Jo, N. Ubrig, H. Berger, A. B. Kuzmenko and A. F. Morpurgo,
Nano Lett. {\bf 14}, 2019 (2014).}

\bibitem{ZhuYe}
{B. Zhu, X. Chen, and X. Cui, Sci. Rep. {\bf 5}, 9218 (2015); Z.
Ye, T. Cao, K. \'{O}Brien, H. Zhu, X. Yin, Y. Wang, S.G. Louie,
and X. Zhang, Nature {\bf 513}, 214 (2014).}

\bibitem{Salehi}
{S. Salehi and A. Saffarzadeh, Sur. Sci. {\bf 651}, 215 (2016).}

\bibitem{Xiao}
{D. Xiao, G-B Liu, W. Feng, X. Xu and W. Yao, Phys. Rev. Lett.
{\bf 108}, 196802 (2012).}

\bibitem{Yao}
{W. Yao, D. Xiao, and Q. Niu, Phys. Rev. B {\bf 77}, 235406
(2008).}

\bibitem{Zeng}
{H. Zeng, J. Dai, W. Yao, D. Xiao, X. Cui, Nat. Nanotech. {\bf 7},
490 (2012).}

\bibitem{Mak}
{K. F. Mak, K. He, J. Shan, and T. F. Heinz, Nat. Nanotech. {\bf
7}, 494 (2012).}

\bibitem{Sie}
{E. J. Sie, J. W. McIver, Y.-H. Lee, L. Fu, J. Kong, and N. Gedik,
Nat. Mat. {\bf 14}, 290 (2015).}

\bibitem{Ye}
{Y. Ye, J. Xiao, H. Wang, Z. Ye, H. Zhu, M. Zhao, Y. Wang, J.
Zhao, X. Yin and X. Zhang,  Nat. Nanotech. {\bf 11}, 598 (2016).}

\bibitem{Wu}
{Z. Wu, F. Zhai, F.M. Peeters, H.Q. Xu, and K. Chang, Phys. Rev.
Lett. {\bf 106}, 176802 (2011).}

\bibitem{Park}
{C.H. Park, Y.W. Son, L. Yang, M.L. Cohen, S.G. Louie, Nano Lett.
{\bf 8}, 2920 (2008).}

\bibitem{Cheianov}
{V. V. Cheianov, V. Fal'ko, and B. L. Altshuler, Science {\bf
315}, 1252 (2007).}

\bibitem{Moghaddam}
{A. G. Moghaddam and M. Zareyan, Phys. Rev. Lett. {\bf 105},
146803 (2010).}

\bibitem{Goos}
{F. Goos and H. H\"{a}nchen, Ann. Phys. (Leipzig) {\bf 436}, 333
(1947).}

\bibitem{Beenakker}
{C.W.J. Beenakker, R.A. Sepkhanov, A.R. Akhmerov, and J.
Tworzydlo, Phys. Rev. Lett. {\bf 102}, 146804 (2009).}

\bibitem{Zhai}
{F. Zhai, Y. Ma, and K. Chang, New J. Phys. {\bf 13}, 083029
(2011).}

\bibitem{Ghosh}
{S. Ghosh, M. Sharma, J. Phys.: Condens. Matter {\bf 21}, 292204
(2009).}

\bibitem{Agrawal}
{N. Agrawal, S. Ghosh, and M. Sharma, Int. J. Mod. Phys. B {\bf
27}, 1341003 (2013).}

\bibitem{Zhang}
{Q. Zhang and K.S. Chan, Appl. Phys. Lett. {\bf 105}, 212408
(2014).}

\bibitem{Azarova}
{E.S. Azarova and G.M. Maksimova, J. Phys. Chem. Sol. {\bf 100},
143 (2017).}

\bibitem{Sun}
{J.F. Sun and F. Cheng, J. Appl. Phys. {\bf 115}, 133703 (2014).}

\bibitem{ChenEPJB}
{X. Chen, J.-W. Tao, and Y. Ban, Eur. Phys. J. B {\bf 79}, 203
(2011).}

\bibitem{ChenJoO}
{X. Chen, X.-J. Lu , Y. Ban and C.-F. Li, J. Opt. {\bf 15} 033001
(2013).}

\bibitem{ChenPLA}
{X. Chen, C.-F. Li, Y. Bana, Phys. Lett. A {\bf 354} 161 (2006).}

\bibitem{Song}
{Y. Song, H.-C. Wu, and Y. Guo, Appl. Phys. Lett. {\bf 100},
253116 (2012).}

\bibitem{ChenEPJB2}
{X. Chen, P.-L. Zhao, X.-J. Lu, L.-G. Wang, Eur. Phys. J. B {\bf
86}, 223 (2013).}

\bibitem{Saffar}
{A. Saffarzadeh and G. Kirczenow, Appl. Phys. Lett. {\bf 102},
173101 (2013).}

\bibitem{PNAS}
{X. Li, T. Cao, Q. Niu, J. Shi, and J. Feng, PNAS {\bf 110}, 3738
(2013).}

\bibitem{error}
{Due to a mistake in a minus sign, the authors of Ref.
\onlinecite{ChenEPJB} could not obtain Eq. (11) in Ref.
\onlinecite{Beenakker}.}

\bibitem{Klein}
{O. Klein, Z. Phys. {\bf 53}, 157 (1929).}

\bibitem{BeenakkerRMP}
{C. W. J. Beenakker, Rev. Mod. Phys. {\bf 80}, 1337 (2008).}

\bibitem{Katsnelson}
{M. I. Katsnelson, K. S. Novoselov, and A. K. Geim, Nat. Phys.
{\bf 2}, 620 (2006).}

\bibitem{ChenAPL2009}
{The transmission gap is an interval of energy where the critical
angle is less than the incident angle, so that the electron wave
function in the gated region becomes evanescent. See, X. Chen,
J.-W. Tao, Appl. Phys. Lett. {\bf 94}, 262102 (2009).}

\end{thebibliography}
\end{document}